\documentclass[11pt]{amsart}
\usepackage{amscd}
\def\ml{\mathcal{C}}
\def\ml1{\mathcal{C}^1}
\def\mlb1{\mathcal{C}_{b}^{1}}
\def\mb{\mathbb{R}}
\def\mbn{\mathbb{R}^n}
\def\g{\gamma}

\def\t{\tau}
\def\b{\beta}
\def\s{\sigma}
\def\a{\alpha}
\def\l{\lambda}
\def\frk{\frak}               

\def\Phi{{\frk n}}
\def\pl{\partial}
\def\opn#1#2{\def#1{\operatorname{#2}}} 
\opn\chara{char} \opn\length{\ell} \opn\pd{pd} \opn\rk{rk}
\opn\projdim{proj\,dim} \opn\injdim{inj\,dim} \opn\rank{rank}
\opn\depth{depth} \opn\grade{grade} \opn\height{height}
\opn\embdim{emb\,dim} \opn\codim{codim}

\opn\Tr{Tr} \opn\bigrank{big\,rank}
\opn\superheight{superheight}\opn\lcm{lcm}
\opn\trdeg{tr\,deg}
\opn\reg{reg} \opn\lreg{lreg} \opn\ini{in} \opn\lpd{lpd}
\opn\size{size}
\opn\div{div} \opn\Div{Div} \opn\cl{cl} \opn\Cl{Cl}
\opn\Spec{Spec} \opn\Supp{Supp} \opn\supp{supp} \opn\Sing{Sing}
\opn\Ass{Ass} \opn\Min{Min}
\opn\Ann{Ann} \opn\Rad{Rad} \opn\Soc{Soc}
\opn\Im{Im} \opn\Ker{Ker} \opn\Coker{Coker} \opn\Am{Am}
\opn\Hom{Hom} \opn\Tor{Tor} \opn\Ext{Ext} \opn\End{End}
\opn\Aut{Aut} \opn\id{id}

\opn\nat{nat}
\opn\pff{pf}
\opn\Pf{Pf} \opn\GL{GL} \opn\SL{SL} \opn\mod{mod} \opn\ord{ord}
\opn\Gin{Gin} \opn\Hilb{Hilb}\opn\sdepth{sdepth}
\opn\aff{aff} \opn\con{conv} \opn\relint{relint} \opn\st{st}
\opn\lk{lk} \opn\cn{cn} \opn\core{core} \opn\vol{vol}
\opn\link{link} \opn\star{star}
\opn\gr{gr}

\def\pot#1#2{#1[\kern-0.28ex[#2]\kern-0.28ex]}

\opn\dirlim{\underrightarrow{\lim}}
\opn\inivlim{\underleftarrow{\lim}}
\def\Implies{\ifmmode\Longrightarrow \else
        \unskip${}\Longrightarrow{}$\ignorespaces\fi}
\def\implies{\ifmmode\Rightarrow \else
        \unskip${}\Rightarrow{}$\ignorespaces\fi}
\def\iff{\ifmmode\Longleftrightarrow \else
        \unskip${}\Longleftrightarrow{}$\ignorespaces\fi}

\let\:=\colon
\newtheorem{Theorem}{THEOREM}[section]
\newtheorem{lemma}[Theorem]{LEMMA}

\newtheorem{Remark}[Theorem]{Remark}

\let\epsilon\varepsilon
\let\phi=\varphi
\let\kappa=\varkappa
\def\qed{\ifhmode\textqed\fi
      \ifmmode\ifinner\quad\qedsymbol\else\dispqed\fi\fi}
\def\textqed{\unskip\nobreak\penalty50
       \hskip2em\hbox{}\nobreak\hfil\qedsymbol
       \parfillskip=0pt \finalhyphendemerits=0}
\def\dispqed{\rlap{\qquad\qedsymbol}}

\opn\dis{dis}
\def\pnt{{\raise0.5mm\hbox{\large\bf.}}}

\opn\Lex{Lex}

\begin{document}
\textwidth=13cm
\title{Hamilton-Jacobi equations with jumps: asymptotic stability}

\author{AMIR MAHMOOD, SAIMA PARVEEN}

\begin{abstract}The asymptotic stability of a global solution satisfying Hamilton-Jacobi equations with jumps will be analyzed in dependence on the strong dissipativity of the jump control function and using orbits of the differentiable flows to describe the corresponding characteristic system.
\end{abstract}

\maketitle
\section*{INTRODUCTION}
Asymptotic behavior of Hamilton-Jacobi (H-J) equations with jumps will be concentrated on revealing the significance of dissipativness property acting in the Hamiltonian $H(x,u,p)=<p,g(x,u)>+L(x,u)$ and
the jump functions $h(x,u)\in\mathbb{R}^m , \, (x,u)\in\mathbb{R}^n\times [-1,1]$. Here are analyzed (H-J) equations of the form
\begin{equation}\label{eq.z}
\partial_t u+ H(x,u,\partial_x u)=0 ,\,\,\,\,\, \,t\in[t_j,t_{j+1}),\,\,\,\,\, x\in B(0,1)\subset \mathbb{R}^n
\end{equation}
where the jumps
$$u(t_j,x)=u(t_{j^-},x)+<h(x,u(t_{j^-},x)),\Delta y(t_j)>,\,\,\,\,\,\,\,\, j=0,1,2,\cdot\cdot\cdot$$
are defined by a sequence $\{t_j\}_{j\geq 0}\uparrow \infty$ and a
piecewise constant process
$\{y(t)=y(t_j)\in\mathbb{R}^m:t\in[t_j,t_{j+1}),\,\,\,\, j\geq 0 ,\,\,\,\,
y(0)=0\}$. The Cauchy method of characteristic systems allows to
construct a global solution
$\{(\widehat{x}(t,\lambda),\widehat{u}(t,\lambda))\in
\mathbb{R}^n\times [-1,1]: t\in[t_j,t_{j+1}), \,\,\,\,\lambda\in
\mathbb{R}^n , \,\,\,\,j\geq 0\}$ provided a weak dissipativity with
respect to $u\in[-1,1]$ of $L$ and $h$ are assumed. On the other
hand, a global bounded solution for the H-J equation \eqref{eq.z} with jumps
is constructed combining $\widehat{u}(t,\lambda)\in[-1,1]$ with a diffeomorphism
 $\{\lambda=\psi(t,.)\in\mathcal{C}^1(D\subset\mathbb{R}^n;\mathbb{R}^n): t\geq 0\}$
 which is piecewise smooth for
 $t\in[t_j,t_{j+1}),\,\,\,\,\, j\geq0$. In doing this we need to
 impose additional conditions which are strictly related with asymptotic stability for the both components of a solution
 $\{u(t,x)=\widehat{u}(t;\psi(t,x)):t\geq 0\}$ and $\{p(t,x)=\partial_x u(t,x): t\geq 0\}$.
 Here, $L(u)\in\mathbb{R},\,\,\,\,h(u)\in\mathbb{R}^m \,\,\,\,\hbox{and}\,\,\,\,u\in[-1,1]$ depend only on unknown solution .
 When  $ g(x,u)=\alpha(u)\widehat{g}(x)\,\,\,\,\,\hbox{and}\,\,\,\,\,\,\widehat{g}\in \mathcal{C}^1(\mathbb{R}^n,\mathbb{R}^n)$
  agrees with a nonlinear growth condition, the analysis of the asymptotic stability is performed in Theorem \ref{th;1},
  but the conclusion will be valid only locally with respect to $x\in D\subset\mathbb{R}^n$. In
section 1 of this paper some auxiliary results regarding to global
existence of the characteristic system solution are given
emphasizing that the both components
   $\{\big(\widehat{u}(t,\lambda),\partial_{\lambda} \widehat{u}(t,\lambda)\big)\in \mathbb{R}^{n+1}: t\geq 0,\,\,\,\, \lambda\in \mathbb{R}^n\}$ are bounded. Here are included two results
   (Lemma \ref{l;3}, Lemma \ref{l;4})
   where the construction of the smooth mapping $\{\lambda=\psi(t,x):t\geq 0,\,\,\,\,x\in D\subseteq\mathbb{R}^n\}$ is given.
    Usually, the construction of the mentioned smooth mapping involves a backward integral equation combined
    with a contractive mapping theorem when the original equation $\widehat{x}(t;\lambda)=x$ can be rewritten as
    $\widehat{G}(\tau(t;\lambda))[\lambda]=x , \,\,\,\,\,\, \tau(t;\lambda)=\int_{0}^{t}\alpha(\widehat{u}(s;\lambda))ds$,
    where $  \{\widehat{G}(\sigma)[\lambda]:\sigma\in\mathbb{R},\,\,\,\, \,\lambda\in\mathbb{R}^n\}$ is the global flow
    generated by $\widehat{g}\in\mathcal{C}_{b}^1(\mathbb{R}^n;\mathbb{R}^n)$. As far as $\widehat{x}(t;\lambda)=x$ and
    $\psi(t,\widehat{x}(t;\lambda))=\lambda$ for any $t\geq 0,\,\,\,\, \,x\in\mathbb{R}^n ,\,\,\,\,\, \lambda\in\mathbb{R}^n$
    we get a direct verification that $\{u(t,x)\mathop{=}\limits^{\hbox{def}}\widehat{u}(t;\psi(t,x)):\,t\geq 0,\,\,\,\,\, x\in\mathbb{R}^n\}$ is the solution of the H-J equation with jumps.
    The situation is changing when the vector field
    $\widehat{g}\in \widehat{\mathcal{C}}^1(\mathbb{R}^n,\mathbb{R}^n)$ is a nonlinear and unbounded one. It is
    analyzed in Theorem \ref{th;1}
    pointing out that the contractive mapping
    theorem can be applied only locally (see $x\in D=\hbox{int} B(x_*,\gamma)\subseteq \mathbb{R}^n$). It implies that a
    nonstandard method must be used in order to get that $\{u(t,x)\mathop{=}\limits^{\hbox{def}}\widehat{u}(t;\psi(t,x)):\,t\geq 0,\,\,\,\,\, x\in D\subseteq\mathbb{R}^n\}$
    is the solution of the H-J
    equations with jumps.
The last theorem in section 2 (see Theorem \ref{th;2}) is telling
us that in the case, when $g(x,u)=\mathop{\sum}\limits_{k=1}^{l}\alpha_{k}(u)\widehat{g}_{k}(x)$
is a summation of several vector fields of the type used in
Theorem \ref{th;1}, we need to impose a commutativity
hypothesis:\,$\{\widehat{g}_1,...,\widehat{g}_{l}\}\subseteq\widehat{\mathcal{C}}^{1}(\mathbb{R}^n,\mathbb{R}^n)$
are commuting using Lie product.
\section{SOME AUXILIARY RESULTS}
We are given an increasing sequence $\{t_j\}_{j\geq 0}$, and a piecewise constant process
$$\{y(t)=y(t_j)\in\mathbb{R}^m:t\in[t_j,t_{j+1}),\,\,\,\, j\geq 0 ,\,\,\,\, y(0)=0\}$$
for which $\Delta y(t_j)=y(t_j)-y(t_{j^-}), \,\,\,\,\, j\geq 1$ can be taken as a bounded value from $\mathbb{R}_{+}^{m}$ and suitable for a fixed goal. Consider the following H-J equation with jumps
\begin{eqnarray}\label{eq:1}\left\{
  \begin{array}{ll}
   \partial_t u(t,x)+&H(x,u(t,x),\partial_x u(t,x))= 0,\,\,t\in[t_j,t_{j+1}),\,x\in\mathbb{R}^n,\, u\in\mathbb{R} \\
    u(t_j,x)= &u(t_{j^{-}},x)+<h(x,u(t_{j^{-}},x)),\Delta y(t_j)>,\,\,j\geq 0\\
    u(0,x)=&u_0(x),\,\,\, u_0\in\widehat{\mathcal{C}}_{b}^1(\mathbb{R}^{n+1})
  \end{array}
\right.
\end{eqnarray}
where $H(x,u,p)=<p,g(x,u)>+L(x,u),\,\,g\in\widehat{\mathcal{C}}_{b}^1(\mathbb{R}^{n+1};\mathbb{R}^n),\,\,\,L\in\mathcal{C}^1(\mathbb{R}^{n+1})$ and $h\in\widehat{\mathcal{C}}_{b}^1(\mathbb{R}^{n+1})$. By $\widehat{\mathcal{C}}_{b}^1$ we denote the space consisting of all continuously differentiable functions $ f(x,u):\mathbb{R}^n\times\mathbb{R}\longmapsto \mathbb{R}^k \,\,(k=n,1,m)$ for which the partial derivatives $\partial_i f(x,u)=\partial_{x_i}f(x,u)$ and $\partial_u f(x,u),\,\,i\in\{1,...,n\}$, are bounded.
To construct a global solution for \eqref{eq:1} which is bounded, we need to convince ourselves that the global solution \begin{equation*}\label{eq.2}
\{(\widehat{x}(t,\lambda),\widehat{u}(t,\lambda))\in\mathbb{R}^{n+1}:\,t\in[t_j,t_{j+1}),\,\,\,\lambda\in\mathbb{R}^{n+1},\,\,j\geq0\} \end{equation*}
exists satisfying the corresponding characteristic system
\begin{eqnarray}\label{eq.3}
\left\{
  \begin{array}{ll}
      \frac{d\widehat{x}}{dt}&=g(\widehat{x},\widehat{u}),\,\,\,\,\,\,\widehat{x}(0,\lambda)=\lambda\in\mathbb{R}^n,\,\,t\geq0\\
        \frac{d\widehat{u}}{dt}&=-L(\widehat{x},\widehat{u}),\,\,t\in[t_j,t_{j+1})\\
        \widehat{u}(t_j;\lambda)&=\widehat{u}(t_{j^{-}};\lambda)+<h(\widehat{x}(t_j,\lambda),\widehat{u}(t_{j^{-}};\lambda)),\Delta y(t_j)>,\,j\geq 0\\
        \widehat{u}(0;\lambda)&=u_0(\lambda),\,\,u_0\in\mathcal{C}_{b}^{1}(\mathbb{R}^n)
  \end{array}
\right.
\end{eqnarray}
 Without any dissipativity conditions on $L$ and $h$, the standard analysis is assuming
 $L\in\widehat{\mathcal{C}}_{b}^{1}(\mathbb{R}^{n+1})$ and
$g\in\widehat{\mathcal{C}}_{b}^{1}(\mathbb{R}^{n+1},\mathbb{R}^n)$,
and we get a unique global solution of \eqref{eq.3}
\begin{equation*}\label{eq.4}
\{z(t,\lambda)=(\widehat{x}(t,\lambda),\widehat{u}(t,\lambda))\in\mathbb{R}^{n+1}:\,t\geq0,\,\,\,\lambda\in\mathbb{R}^n\}
\end{equation*}
On each interval $t\in[t_j,t_{j+1})$ we obtain a
smooth function
$\{z(t;\l)\in\mb^{n+1},\,\,t\in[t_j,t_{j+1}),\,\,\l\in\mb^n\}$
satisfying
 \begin{equation*}\label{eq.5}
 \frac{dz}{dt}=Z(z), \,\,t\in[t_j,t_{j+1}),\,\,z(t_j;\l)=z(t_{j^{-}};\lambda)+b(z(t_{j^{-}};\l),\Delta y(t_j)),\,j\geq0
 \end{equation*}
 where
$z(0;\lambda)=(\lambda,u_0(\lambda))=z_0(\lambda)\in\mathbb{R}^{n+1}$ and
\begin{eqnarray*}\label{eq.6}
\left\{
  \begin{array}{ll}
              Z(z)&\mathop{=}\limits^{\hbox{def}}\left(
                                                    \begin{array}{c}
                                                      g(\widehat{x},\widehat{u}) \\
                                                      -L(\widehat{x},\widehat{u}) \\
                                                    \end{array}
                                                  \right)
              \in\mathbb{R}^{n+1},\,\,z=(\widehat{x},\widehat{u})\in\mathbb{R}^{n+1}\\
       b(z,\nu)&=\left(
                  \begin{array}{c}
                    0 \\
                    <h(z),\nu> \\
                  \end{array}
                \right),\,\,\hbox{for null element}\,\,0\in\mathbb{R}^n
  \end{array}
\right.
\end{eqnarray*}
Using the piecewise smooth solution
$z(t,\lambda)=(\widehat{x}(t,\lambda),\widehat{u}(t,\lambda)),\,t\in[t_j,t_{j+1}),\,\,\lambda\in\mathbb{R}^n,\,\,j\geq0$
of \eqref{eq.3} we get the solution of the H-J equation \eqref{eq:1} as a composition
\begin{equation}\label{eq.7}
u(t,x)\mathop{=}\limits^{\hbox{def}}\widehat{u}(t,\psi(t,x)),\,\,t\geq0,\,\,x\in\mathbb{R}^n
\end{equation}
where the smooth mapping $\{\lambda=\psi(t,x)\in\mathbb{R}^n:\,t\geq0,\,\,x\in\mathbb{R}^n\}$
is found as a unique solution fulfilling
\begin{equation*}\label{eq.8}
\widehat{x}(t;\lambda)=x\,\in\,\mathbb{R}^n,\,\,t\geq0
\end{equation*}
\begin{equation}\label{eq.9}
\psi(0,x)=x,\,\,\widehat{x}(t,\psi(t,x))=x,\,\,(\forall)\,t\geq0
\end{equation}
As far as $\{\lambda=\psi(t,x)\in\mathbb{R}^n:\,t\geq0,\,\,x\in\mathbb{R}^n\}$ is the unique solution of the equation \eqref{eq.9}, we notice that
\begin{equation}\label{eq.10}
\psi(t,\widehat{x}(t;\lambda))=\lambda,\,\,\,\lambda\,\in\,\mathbb{R}^n,\,\,t\geq0
\end{equation}
and this will be the main ingredient supporting the idea that
$\{u(t,x):\,t\in[t_j,t_{j+1}),\,\,x\in\mathbb{R}^n,\,\,j\geq0\}$
defined in \eqref{eq.7} is the solution of H-J equation
\eqref{eq:1}. The construction of the mapping
$\{\lambda=\psi(t,x)\in\mathbb{R}^n:\,t\geq0,\,\,x\in\mathbb{R}^n\}$
is given using bounded solution
$\{\widehat{u}(t;\lambda):\,t\geq0,\,\,\lambda\in\mathbb{R}^n\}$
for which the gradient
$\{\pl_\l\widehat{u}(t,x):\,t\geq0,\,\,\lambda\in\mathbb{R}^n\}$
is a bounded mapping. In this respect we recall that
$\{\widehat{u}(t,x)\in\mathbb{R}:\,y\in[t_j,t_{j+1}),\,\,j\geq0,\,\,\lambda\in\mathbb{R}^n\}$
is a piecewise smooth scalar function and
$\{\widehat{x}(t,\l)\in\mb^n:\,t\geq0,\,\,\l\in\mb ^n\}$ is a
smooth mapping satisfying
\begin{equation*}\label{eq.11}
\widehat{x}(t,\lambda)=\widehat{x}(t_j,\lambda)+\int_{t_j}^{t}g(\widehat{x}(s,\lambda),\widehat{u}(s,\lambda))ds,\,t\in[t_j,t_{j+1}),\,\,j\geq1
\end{equation*}

\begin{equation*}\label{eq.12}
\widehat{x}(t,\lambda)=\lambda+\int_{0}^{t}g(\widehat{x}(s,\lambda),\widehat{u}(s,\lambda))ds,\,t\in[0,t_1]
\end{equation*}
\begin{eqnarray}\label{eq.13}
\left\{
  \begin{array}{ll}
   \frac{d\widehat{u}}{dt}&= -L(\widehat{x}(t;\lambda),\widehat{u}) ,\,\,\,t\in[t_j,t_{j+1})\\
    \widehat{u}(t_j;\lambda)&= \widehat{u}(t_{j^{-}};\lambda)+ <h(\widehat{x}(t_j;\lambda),\widehat{u}(t_{j^{-}};\lambda)),\Delta y(t_j)>\, , \,\,j\geq 1\\
    \widehat{u}(0;\lambda)&=u_0(\lambda),\,\,\, u_0\in\widehat{\mathcal{C}}_{b}^1(\mathbb{R}^{n}) ,\,\,sup\mid u_0(x)\mid=k_0< 1
  \end{array}
\right.
\end{eqnarray}
The following weak dissipativity conditions will be assumed
\begin{eqnarray}\label{eq.14}
\left\{
  \begin{array}{ll}
   L(x,0)= &0 ,\,\,\,\partial_u L(x,u)\geq0,\,\,(x,u)\in\mathbb{R}^n\times[-1,1]\\
h_i(x,0)=&0 ,\,\,\, \partial_u h_i(x,u)\leq0,\,\,\,i\in\{1,...,m\},\,\,(x,u)\in\mathbb{R}^n\times[-1,1]
  \end{array}
\right.
\end{eqnarray}
\begin{lemma}\label{l;1}
Consider
$g\in\widehat{\mathcal{C}}_{b}^1(\mathbb{R}^{n+1};\mathbb{R}^n)$
and let $L\in\mathcal{C}^1(\mathbb{R}^{n+1})$ and
$h\in\widehat{\mathcal{C}}_{b}^1(\mathbb{R}^n\times[-1,1];\mathbb{R}^m)$
be fulfilling the condition \eqref{eq.14}. Let $\Delta
y(t_j)\in\mathbb{R}_{+}^m$ be such that
\begin{equation*}
-1\leq\big<\partial_u h(x,u),\Delta y(t_j)\big>\leq 0
,\,\,(x,u)\in\mathbb{R}^n\times[-1,1], \,\,\,j\geq 1
\end{equation*}
Then the unique global solution $\{\big(\widehat{x}(t;\lambda),\widehat{u}(t;\lambda)\big): t\geq 0,\,\lambda\in\mathbb{R}^n\}$ of the characteristic system \eqref{eq.13}
satisfies
\begin{eqnarray*}\left\{
  \begin{array}{ll}
   &\mid \widehat{u}(t_j;\lambda)\mid \leq\mid \widehat{u}(t_{j^{-}};\lambda)\mid ,\,\,\mid \widehat{u}(t;\lambda)\mid \leq\mid \widehat{u}(t_{j^{-}};\lambda)\mid ,\,t\in[t_j,t_{j+1}),\,j\geq 1 \\
    &\mid \widehat{u}(t;\lambda)\mid\leq\mid u_0(\lambda)\mid\leq 1,\,\,\forall \,t\geq 0,\,\,\lambda\in\mathbb{R}^n
  \end{array}
\right.
\end{eqnarray*}
\end{lemma}
\begin{Remark}\label{r.1}
Taking $u_0\in\mathcal{C}_{b}^{1}(\mathbb{R}^n)$ such that $sup\mid u_0(x)\mid=K_0<1$, under the hypothesis of Lemma \ref{l;1}, we get a solution $\{\widehat{u}(t;\lambda):\,t\geq0,\,\,\lambda\in\mathbb{R}^n\}$ of \eqref{eq.13} fulfilling $\mid\widehat{u}(t;\lambda)\mid\leq1,\,\,\,\,t\geq0,\,\,\lambda\in\mathbb{R}^n$. Now we are interested to obtain a bounded solution for which $\{\pl_\l\widehat{u}(t,\l):\,t\geq0,\,\,\l\in\mb^n\}$ is bounded and $\mid\pl_\l u(t;\l)\mid\leq\mid\pl_\l u_0(t;\l)\mid\,,\,\,t\geq0,\,\,\l\in\mb^n$. In this respect, we assume
\begin{eqnarray}\label{eq.21}
 \left\{
  \begin{array}{ll}
   (a)\,&L\in\mathcal{C}^1(\mathbb{R}),\, L(0)=0,\,\,\partial_u L(u)\geq 0,\,\,u\in[-1,1];\\
    (b)\,&h_i\in\mathcal{C}^1([-1,1]),\,h_i(0)=0,\partial_uh_i(u)\leq 0,\,\,u\in[-1,1],\,\,i\in\{1,...,m\};\\
   (c)\, &g\in\widehat{\mathcal{C}}_{b}^{1}(\mathbb{R}^n\times[-1,1];\mathbb{R}^n)\,\,\hbox{and}\,\,u_0\in\mathcal{C}_{b}^{1}(\mathbb{R}^n) \,\,\hbox{satisfies}\\
   &\,\,sup\mid u_0(x)\mid=K_0<1
  \end{array}
\right.
\end{eqnarray}
\end{Remark}
\begin{lemma}\label{l;2}
Consider $H(x,u,p)=<p,g(x,u)>+L(u)$, where $g$ and $L$ fulfil \eqref{eq.21}. Let $h(u)\in\mathbb{R}^m$ and $u_0\in\mathcal{C}_{b}^{1}(\mathbb{R})$ be such that \eqref{eq.21} are satisfied. Take $\Delta y(t_j)\in\mathbb{R}_{+}^{m}$ such that
\begin{equation*}\label{eq.22}
0\geq\,<\partial_u h(u),\Delta y(t_j)>\,\geq -1,\,\,u\in[-1,1],\,\,j\geq 1,
\end{equation*}
and consider the global solution
\begin{equation*}\label{eq.23}
\{\widehat{u}(t;\lambda):\,t\geq0,\,\,\lambda\in\mathbb{R}^n\}
\end{equation*}
satisfying \eqref{eq.3}. Then, it holds
\begin{eqnarray*}\label{eq.24}
\left\{
  \begin{array}{ll}
   \mid\widehat{u}(t;\lambda)\mid\leq\mid u_0(\lambda)\mid\leq 1,\,\,\,t\geq0,\,\,\lambda\in\mathbb{R}^n \\
    \mid\partial_\lambda\widehat{u}(t;\lambda)\mid\leq\mid \partial_\lambda u_0(\lambda)\mid,\,\,\,t\geq0,\,\,\lambda\in\mathbb{R}^n \\
\end{array}
\right.
\end{eqnarray*}
\end{lemma}
\begin{Remark}
The two previously given lemmas help us to conclude that looking
for a bounded global solution
$\{u(t;x):\,t\geq0,\,\,x\in\mathbb{R}^n\}$ for which the gradient
$\{p(t,x)=\partial_xu(t;x):\,t\geq0,\,\,x\in\mathbb{R}^n\}$ is
also a bounded function we need to assume some weak dissipativity
condition for both $L\in\mathcal{C}^1(\mb)$ and
$h\in\mathcal{C}^1([-1,1];\mb^m)$. Unfortunately, the same
dissipativity conditions are not sufficient for obtaining the
smooth mapping $\{\l=\psi(t,x)\in\mb^n:\,t\geq0,\,\,x\in\mb^n\}$
satisfying equations \eqref{eq.9} and \eqref{eq.10}. We shall
present two types of necessary conditions which lead us to the
smooth global mapping $\{\l=\psi(t,x):t\geq0, \,\,x\in\mb^n\}$.
For $L$ and $g$ we suppose
\begin{eqnarray}\label{eq.32}
\left\{
  \begin{array}{ll}
   (a)&\,\,\, L\in\mathcal{C}^1(\mathbb{R}),\, L(0)=0,\,\,0<\gamma\leq\partial_u L(u),\,\,u\in[-1,1]\\
    (b)&\,\,\,g(x,u)=\alpha(u)\widehat{g}(x),\,\,\alpha\in\mathcal{C}^1[-1,1]),\,\,\,\widehat{g}\in\mathcal{C}_{b}^{1}(\mb^n,\mb^n),\,\,\alpha(0)=0
   \end{array}
\right.
\end{eqnarray}
Denote $C_1=max\{\mid\pl_u\alpha(u)\mid:\,\,\,u\in[-1,1]\},\,\,\,C_2=sup\{\mid\widehat{g}(x)\mid:\,\,\,x\in\mb^n\}$
and assume that $u_0$ and $h$ fulfil
\begin{eqnarray}\label{eq.33}
\left\{
  \begin{array}{ll}
   (a)&\,\, u_0\in\mathcal{C}_{b}^{1}(\mathbb{R}^n),\, \,sup\mid u_0(x)\mid=K_0<1,\,\, sup\mid \pl_x u_0(x)\mid=K_1;\\
    (b)&\,\,\frac{1}{\gamma}C_1C_2K_1\leq\rho,\,\,\hbox{where}\,\,\rho\in(0,1)\,\hbox {is fixed};\\
   (c)&\,\,h_i\in\ml1([-1,1]),\,\,\,h_i(0)=0,\,\,\,\pl_uh_i(u)\leq 0,\,\,\,u\in[-1,1],\,\,\,i\in\{1,...,m\}
  \end{array}
\right.
\end{eqnarray}
\end{Remark}
 \begin{lemma}\label{l;3}
Consider $H(x,u,p)=<p,g(x,u)>+L(u)$ and the corresponding characteristic system \eqref{eq.3} where $g,L,u_0$ and $h_i,\,\,i\in\{1,...,m\}$, fulfil \eqref{eq.32} and \eqref{eq.33}. Take $\Delta y(t_j)\in\mb_{+}^{m}$ sufficiently small verifying
\begin{equation*}\label{eq.34}
0\geq <\pl_uh(u),\Delta y(t_j)>\,\geq -1,\,\,\,u\in[-1,1],\,\,j\geq 1
\end{equation*}
Then there exist a unique global solution $\{(\widehat{x}(t,\l),\widehat{u}(t,\l)):\,\,t\ge0,\,\l\in\mb^n\}$ of \eqref{eq.3} and a smooth mapping $\{\l=\psi(t,x):\,t\geq0,\,x\in\mb^n\}$ fulfilling
\begin{equation}\label{eq.35}
\widehat{x}(t;\psi(t,x)=x,\,\,\psi(t,\widehat{x}(t,\l))=\l,\,\psi(0,x)=x ,\,      \,t\geq0,\,\,x,\lambda\in\mathbb{R}^n,
\end{equation}
where $\mid\widehat{u}(t,\l)\mid\leq K_0<1$ and
$\mid\pl_\l\widehat{u}(t,\l)\mid\leq
K_1,\,\,\,t\geq0\,,\,\,\l\in\mb^n$
\end{lemma}
\begin{Remark}
The unique global solution $\{\l=\psi(t,x):\,t\geq0,\,x\in\mb^n\}$ given in Lemma \ref{l;3} fulfils the equation \eqref{eq.35}, $\psi_i(t,\widehat{x}(t,\l))=\l_i,\,\,t\geq0,\,\,i\in\{1,...,n\}$, where $\l=(\l_1,...,\l_n)\in\mb^n$. By a direct computation we get
\begin{equation}\label{eq.57}
\frac{d}{dt}[\psi_i(t,\widehat{x}(t;\l))]=\pl_t\psi_i(t,\widehat{x}(t;\l))+<\pl_x\psi(t,\widehat{x}(t;\l)),g(\widehat{x}(t,\l),\widehat{u}(t;\l))>=0
\end{equation}
for any $t\geq 0,\,i\in\{1,...,n\}$ and $\l\in\mb^n$. In addition for $\l=\psi(t,x)$ we obtain $\widehat{x}(t,\psi(t,x))=x,\,\,\widehat{u}(t,\psi(t,x))=u(t,x)$ and the equation \eqref{eq.57} becomes $\psi_i(0,x)=x_i$ and
\begin{equation}\label{eq.58}
\pl_t\psi_i(t,x)+<\pl_x\psi_i(t,x),g(x,u(t,x))>=0,\,\,(\forall)\,\,t\geq0,\,x\in\mb^n,\,\,i\in\{1,...,n\}
\end{equation}
where $g(x,u)=\alpha(u)\widehat{g}(x)$. We are going to analyze a second type of conditions which leads us to a smooth mapping $\{\l=\psi(t,x):\,\,t\geq0,\,x\in\mb^n\}$ satisfying \eqref{eq.58} and a solution $u(t,x)=\widehat{u}(t,\psi(t,x)),\,t\geq0,\,\,x\in\mb^n$ of H-J equation \eqref{eq:1} which is only asymptotically stable. In this respect, suppose $L$ and $g$ fulfil
 \begin{eqnarray}\label{eq.59}
\left\{
  \begin{array}{ll}
   (a)&\,\,\, L\in\mathcal{C}^1(\mathbb{R}),\, L(0)=0,\,\,\pl_uL(u)\geq0,\,\,u\in[-1,1]\\
    (b)&\,\,\,g(x,u)=\alpha(u)\widehat{g}(x),\,\,\alpha\in\mathcal{C}^1[-1,1]),\,\,\,\widehat{g}\in\mathcal{C}_{b}^{1}(\mb^n,\mb^n),\,\,\alpha(0)=0
   \end{array}
\right.
\end{eqnarray}
Denote $C_1=max\{\mid\pl_u\alpha(u)\mid:\,\,\,u\in[-1,1]\},\,\,\,C_2=sup\{\mid\widehat{g}(x)\mid:\,\,\,x\in\mb^n\}$
and take $u_0$ and $h$ such that
\begin{eqnarray}\label{eq.60}
\left\{
  \begin{array}{ll}
   (a)\,\,\, &u_0\in\mathcal{C}_{b}^{1}(\mathbb{R}^n),\, \,sup\mid u_0(x)\mid=K_0,\,\, sup\mid \pl_u u_0(x)\mid=K_1\\
   \\
    (b)\,\,\,&h_i\in\ml1([-1,1]),\,h_i(0)=0,\,\pl_uh_i(u)\leq0,\,i\in\{1,...,m\}\\
    &\hbox{and}\,\,\mathop{\sum}\limits_{i=1}^{m}\pl_uh_i(u)\leq-\delta<0,\,\,u\in[-1,1]\\
   (c)\,\,\, &2dC_1C_2K_1\leq\rho\,\,\,\hbox{where}\,\,\rho\in(0,1)\,\hbox{is fixed and}\,d=\mathop{max}\limits_{j\geq0}(t_{j+1}-t_j)
  \end{array}
\right.
\end{eqnarray}
\end{Remark}
\begin{lemma}\label{l;4}
Consider $H(x,u,p)=<p,g(x,u)>+L(u)$, where $g$ and $L$ fulfil \eqref{eq.59}. Let $u_0$ and $h_i ,\,\,i\in\{1,...,m\}$, be such that the conditions \eqref{eq.60} are satisfied. Take $\Delta y(t_j)\in\mb_{+}^{m}$ satisfying the inequalities
\begin{equation*}\label{eq.61}
-1\leq<\pl_uh(u),\Delta y(t_j)>\,\leq -\frac{1}{2},\,u\in[-1,1],\,\,j\geq1
\end{equation*}
Then a smooth mapping $\{\l-\psi(t,x):\,t\geq0,\,x\in\mb^n\}$ exists satisfying the equations
\begin{equation*}\label{eq.62}
\widehat{x}(t,\psi(t,x))=x,\,\,\psi(0,x)=x,\,\psi(t,\widehat{x}(t,\l))=\l,\,t\geq0,\,\,x,\l\in\mb^n
\end{equation*}
where $\{(\widehat{x}(t;\l),\widehat{u}(t;\l)):\,t\geq0,\,\l\in\mb^n\}$ is the global solution of \eqref{eq.3} and $\mid\widehat{u}(t,\l)\mid\leq 1,\,\,\,t\geq0,\,\l\in\mb^n$

\end{lemma}

\section{MAIN RESULTS}
The first theorem concern the local asymptotic stability when the
vector field $g(x,u)=\alpha(u)\widehat{g}(x)$ and
$\widehat{g}\in\mathcal{C}^1(\mathbb{R}^{n};\mathbb{R}^n)$
agrees with a nonlinear growth condition. The second theorem given
here analysis the asymptotic stability property when the vector
field
$g(x,u)=\alpha_1(u)\widehat{g}_1(x)+\alpha_2(u)\widehat{g}_2(x)$,
contains two commuting vector fields
$\widehat{g}_1,\widehat{g}_2\in\mathcal{C}^1(\mathbb{R}^{n};\mathbb{R}^n)$.
Asymptotic stability of H-J equation
$$\partial_t u+H(x,u,\partial_x u)=0$$
is concerned. It will be the goal of the next two theorems. We are
going to construct a local asymptotically stable solution for
H-J equation \eqref{eq:1} and in this respect we need to
assume (for some $u_*\in\mathbb{R},\,x_*\in\mathbb{R}^n$ fixed)
\begin{eqnarray}\label{eq.104}
\left\{
  \begin{array}{ll}
   (a)&\,\,\, L\in\mathcal{C}^1(\mathbb{R}),\, L(u_*)=0,\,\,\partial_u L(u)\geq 0,\,\,u\in[a,b]\\
    (b)&\,\,\,g(x,u)=\alpha(u)\widehat{g}(x),\,\,\alpha\in\mathcal{C}^1[a,b]),\,\alpha(u_*)=0\\
   (c)&\,\,\,\widehat{g}\in{\mathcal{C}}^1(\mathbb{R}^{n};\mathbb{R}^n), \,\,\widehat{g}(x_*)=0
  \end{array}
\right.
\end{eqnarray}
where $a=u_*-1,\,\,b=u_*+1$. Denote\\
$c_1=max\{\mid\partial_x\alpha(u)\mid:
u\in[a,b]\},\,\,\,c_2=\max\{\mid\widehat{g}(x)\mid:\,x\in\overline{B(x_*,\gamma)}\}$
where $\gamma>0$ is fixed. Take $u_0$ and $h$ such that
\begin{eqnarray}\label{eq.105}
\left\{
  \begin{array}{ll}
   (a)&u_0\in\mathcal{C}_{b}^1(\mathbb{R}^n),\,\,sup\mid u_0(x)\mid=K_0<1,\,sup\mid\partial_x u_0(x)\mid=K_1;\\
    (b)&h_i\in\mathcal{C}^1([a,b]),\,\,h_i(u_*)=0,\,\,\partial h_i(u)\leq 0 \\ &\hbox{and}\,\mathop{\sum}\limits_{i=1}^{m}\partial_u h_i(u)\leq\delta< 0,
    i\in\{1,...,m\},\,u\in[a,b];\\
   (c)&2dc_1c_2K_1\leq\rho,\,\hbox{where}\, \rho\in(0,1)\,\,\hbox{and}\,\,
   d=\mathop{max}\limits_{j\geq0}(t_{j+1}-t_j)\\
   &\hbox{satisfies}\,\,2dc_1c_2\leq\gamma
  \end{array}
\right.
\end{eqnarray}
\begin{Theorem}\label{th;1}
Consider $H(x,u,p)=<p,g(x,u)>+L(u)$, where $g,L$ satisfy \eqref{eq.104} and let $u_0, \,h$ be such that \eqref{eq.105} are verified. Take $\Delta y(t_j)\in\mathbb{R}_{+}^{m}$ fulfilling\\
\begin{equation*}\label{eq.106}
-1\leq\,<\partial_u h(u),\Delta y(t_j)>\,\leq -\frac{1}{2},\,\,\,u\in[a,b],\,\,j\geq 0
\end{equation*}
Then there exists an asymptotic stable solution\\
$$\{u(t,x)\in[a,b]:\,t\in[t_j,t_{j+1}),\,x\in B(x_*,\gamma),\,j\geq 0\}$$
of H-J equations with jumps \eqref{eq:1} verifying  $
u(0,x)=u_*+u_0(x)$
$$\mid u(t,x)-u_*\mid\leq(\frac{1}{2})^{j},\,\,\,t\in[t_j,t_{j+1}),\,\,x\in B(x_*,\gamma),\,\,j\geq 0$$
$$\mid \partial_i u(t,x)\mid\,\leq \frac{L_iK_1}{1-\rho}(\frac{1}{2})^{j},\,\,t\in[t_j,t_{j+1}),\,\,x\in B(x_*,\gamma),\,j\geq 0,\,\,i\in\{1,...,n\}$$
for some constant $L_i>0$
\end{Theorem}
\begin{proof}
By hypothesis, the conclusion of Lemma \ref{l;2} is fulfilled.
Using \eqref{eq.105} and Lemma \ref{r.1},
by a direct computation we obtain the following estimates\\
\begin{eqnarray*}\label{eq.107}
\left\{
  \begin{array}{ll}
\mid\widehat{u}(t;\lambda)-u_*\mid\,\leq \mid u_0(\lambda)\mid\,\leq K_0\,< 1,\\
\,\mid\partial_{\lambda}\widehat{u}(t,\lambda)\mid\,\leq\mid\partial_{\lambda}u_0(\lambda)\mid,\,\,t\geq0,\,\lambda\in\mathbb{R}^n
\end{array}
\right.
\end{eqnarray*}
\begin{equation*}\label{eq.108}
\mid\widehat{u}(t;\lambda)-u_*\mid\,\leq \mid\widehat{u}(t_j;\l)-u_*\mid\leq(\frac{1}{2})^{j}K_0,\,\,t\in[t_j,t_{j+1}),\,\lambda\in\mathbb{R}^n,\,\,j\geq0
\end{equation*}
\begin{equation}\label{eq;109}
\mid\pl_\l\widehat{u}(t;\lambda)\mid\,\leq\,\mid\pl_\l
\widehat{u}(t_j;\lambda)\mid\, \leq(\frac{1}{2})^{j}K_1
,\,\,t\in[t_j,t_{j+1}),\,\lambda\in\mathbb{R}^n,\,\,j\geq0
\end{equation}
This time, the smooth mapping $\{\lambda=\psi(t,x)\}$ will be
found as a unique solution of the integral equation
\begin{equation}\label{eq..21}
\lambda=\widehat{G}(-\tau(t;\lambda))[x],\,\,\tau(t;\lambda)=\mathop{\int}\limits_{0}^{t}\alpha\big(\widehat{u}(s;\lambda)\big)ds,
\,\,\,t\geq 0,\,\,x\in B(x_*,\gamma)\subseteq\mathbb{R}^n
\end{equation}
This time, the vector field $\widehat{g}\in \ml1(\mbn;\mbn)$ is a nonlinear unbounded one and the corresponding local flow $\{\widehat{G}(\s)[x]:\s\in[-\b,\b],\,x\in B(x_*,\g)\}$ is constructed for $\b=2dC_1$ satisfying
\begin{equation}\label{eq..22}
\b c_2\leq\g\,\,\,\,\,(\hbox{see}(\ref{eq.105},c))
\end{equation}
A direct computation shows that
\begin{eqnarray*}\label{eq..23}
|\tau(t,\l)|\leq\mathop{\sum}_{j=0}^{\infty}\mathop{\int}_{t_{j}}^{t_{j+1}}[\mathop{\int}_{0}^{1}|\pl_u\a\big(u_*+\theta(\widehat{u}(t;\l)-u_*)\big)d\theta|]|\widehat{u}(t,\l)-u_*|dt\\ \nonumber
\leq dC_1\mathop{\sum}_{j=0}^{\infty}(\frac{1}{2})^j=2dC_1=\b,\,\,\,\,\,\forall t\geq0,\l\in\mbn
\end{eqnarray*}
Using \eqref{eq..22} we get easily that the right hand side in \eqref{eq..21}
\begin{equation}\label{eq..24}
V(t,x;\l)\mathop{=}^{\hbox{def}}\widehat{G}(-\t(t;\l))[x]=x-[\int_{0}^1\widehat{g}\big(\widehat{G}(-\theta\t(t;\l))\big)[x]d\theta]\t(t;\l)
\end{equation}
satisfies
\begin{equation}\label{eq..25}
|V(t,x;\l)|\leq|x|+2dC_1C_2=|x|+\b C_2\leq 2\g
\end{equation}
for any $t\geq0,\,\,\,x\in B(x_*,\g)$ and $\l\in\mbn$. Here we used that the local flow $\{\widehat{G}(\s)[x]:\,\s\in[-\b,\b],\,\,\,x\in B(x_*,\g)\}$ is bounded fulfilling
\begin{equation*}\label{eq..26}
\widehat{G}(\s)[x]\in B(x_*,2\g),\,\,\hbox{for any} \,\s\in[-\b,\b],\,\,\,x\in B(x_*,\g)
\end{equation*}
It can be seen easily, noticing that the $\{y_k(\s,x):\s\in[-\b,\b],\,\,x\in B(x_*,\g)\}_{k\geq0}$ defining the local flow is uniformly bounded and verifies
\begin{equation*}\label{eq..27}
y_k(\s,x)\in B(x_*,2\g),\,\,\s\in[-\b.\b],\,\,x\in B(x_*,\g),\,\,k\geq0
\end{equation*}
With these notations, the integral equation \eqref{eq..21} can be written as follows
\begin{equation*}\label{eq..26}
\l=V(t,x;\l)
\end{equation*}
where the smooth mapping $V(t,x;\l)$ is a contractive application with respect to $\l\in\mbn$. In this respect, compute
\begin{equation*}\label{eq..29}
M(t,x;\l)=\pl_\l V(t,x;\l)=\widehat{g}(V(t,x;\l))\pl_\l\t(t;\l)
\end{equation*}
where (see \eqref{eq..21})
\begin{equation*}\label{eq..30}
\pl_\l\t(t,\l)=\int_{0}^{t}\pl_u\a\big(\widehat{u}(s;\l)\big)\pl_\l\widehat{u}(s;\l)ds
\end{equation*}
Using \eqref{eq;109} we get
\begin{equation}\label{eq..31}
|\pl_\l\t(t,\l)|\leq C_1K_1\mathop{\sum}\limits_{j=0}^{\infty}(\frac{1}{2})^j=2dC_1K_1
\end{equation}
and from \eqref{eq..25} and \eqref{eq..31} we obtain
\begin{equation*}\label{eq..32}
|M(t,x;\l)|\leq2dC_1C_2K_1=\rho\in(0,1),\,\,\hbox{for any}\,t\geq0,\,x\in B(x_*,\g)\,\hbox{and}\,\l\in\mbn
\end{equation*}
As a consequence, applying the contractive mapping theorem we obtain a smooth mapping
$\{\l=\psi(t,x)\in B(x_*,2\g):t\geq0,x\in B(x_*,\g)\}$
as the unique solution of the integral equation
\begin{equation}\label{eq..33}
\psi(t,x)=V(t,x;\psi(t,x)),\,\,t\geq0,\,x\in B(x_*,\g)
\end{equation}
Define
\begin{equation}\label{eq..34}
u(t,x)=\widehat{u}(t,\psi(t,x)),\,t\geq0,\,x\in B(x_*,\g)
\end{equation}
and to prove that $\{u(t,x):t\geq0,\,x\in B(x_*,\g)\}$ defined in \eqref{eq..34} is the solution for H-J equation \eqref{eq:1} we need to show that the smooth function satisfying \eqref{eq..33} is the solution for the following H-J equations
\begin{equation}\label{eq..35}
\left\{
  \begin{array}{ll}
   \pl_t\psi(t,x)&+\pl_x\psi(t,x)g(x,u(t,x))=0,\,\,t\geq0,\,x\in \hbox{int} B(x_*,\g)  \\
   \psi(0,x)&=x
  \end{array}
\right.
\end{equation}
In this respect, for $\widetilde{t}\in[t_j,t_{j+1})$ and $\widetilde{x}\in \hbox{int} B(x_*,\g)$ fixed define $\{\widehat{x}(s;\widetilde{x}),\widehat{u}(s;\widetilde{x}):\,s\in[\widetilde{t},\widetilde{t}+\varepsilon]\subseteq[t_j,t_{j+1})\}$ as the solution of the characteristic system
\begin{equation}\label{eq..36}
\left\{
  \begin{array}{ll}
    \frac{d\widehat{x}}{ds}=&g(\widehat{x},\widehat{u}),\,\,\,\,\widehat{x}(\widetilde{t};\widetilde{x})=\widetilde{x}\\
   \frac{ d\widehat{u}}{ds}=&-L(\widehat{u}),\,\,\,\widehat{u}(\widetilde{t},\widetilde{x})=\widehat{u}(\widetilde{t};\psi(\widetilde{t},\widetilde{x}))=u(\widetilde{t},\widetilde{x})
  \end{array}
\right.
\end{equation}
satisfying $\widehat{x}(s,\widetilde{x})\in B(x_*,\g),\,\,\,s\in[\widetilde{t},\widetilde{t}+\varepsilon)$. Notice that the solution of \eqref{eq..36} can be obtained as the restriction to $[\widetilde{t},\widetilde{t}+\varepsilon)$ of the global solution $\{\widehat{x}(s;\l),\widehat{u}(s;\l):\,s\geq0\}$ satisfying the original characteristic system
\begin{equation*}\label{eq..37}
\left\{
  \begin{array}{ll}
    \frac{d\widehat{x}}{ds}=&g(\widehat{x},\widehat{u}),\,\,\,\,\widehat{x}(0;\l)=\l\\
   \frac{d\widehat{u}}{ds}=&-L(\widehat{u}),\,\,\,\widehat{u}(t_j,\l)=\widehat{u}(t_{j^-};\l)+<h(\widehat{u}(t_{j^-};\l)),\Delta y(t_j)>,\\
   &s\in[t_j,t_{j+1}),\,j\geq0
  \end{array}
\right.
\end{equation*}
$\widehat{u}(0;\l)=u_0(\l)+u_*$ for $\l=\psi(\widetilde{t},\widetilde{x})$ fixed. As a consequence, we use the representation
\begin{equation*}\label{eq..38}
\widehat{x}(s,\widetilde{x})=\widehat{G}(\t(s;\psi(\widetilde{t},\widetilde{x})))[\psi(\widetilde{t},\widetilde{x})],\,\,s\in[\widetilde{t},\widetilde{t}+\varepsilon]
\end{equation*}
and replacing $x$ by $\{\widehat{x}(s,\widetilde{x}):\,s\in[\widetilde{t},\widetilde{t}+\varepsilon)\}$ into \eqref{eq..24} we get
\begin{eqnarray}\label{eq..39}
V(s,\widehat{x}(s,\widetilde{x});\psi(\widetilde{t},\widetilde{x}))&=\widehat{G}(-\t(s;\psi(\widetilde{t},\widetilde{x})))\circ \widehat{G}(-\t(s;\psi(\widetilde{t},\widetilde{x})))[\psi(\widetilde{t},\widetilde{x})]\\ \nonumber
&=\psi(\widetilde{t},\widetilde{x})=\hbox{const},\,\,s\in[\widetilde{t},\widetilde{t}+\varepsilon)
\end{eqnarray}
Taking derivative with respect to $s$, from \eqref{eq..39}, we obtain the following H-J equations at $s=\widetilde{t}$
\begin{equation}\label{eq..40}
\pl_tV(\widetilde{t},\widetilde{x};\psi(\widetilde{t},\widetilde{x}))+\pl_xV(\widetilde{t},\widetilde{x};\psi(\widetilde{t},\widetilde{x}))g(\widetilde{x},u(\widetilde{t},\widetilde{x}))=0
\end{equation}
where $(\widetilde{t},\widetilde{x})\in[t_j,t_{j+1})\times \hbox{int} B(x_*,\g)$ was arbitrarily fixed. On the other hand, taking derivatives with respect to $t$ and $x$, from \eqref{eq..33} we obtain
\begin{equation}\label{eq..41}
\left\{
  \begin{array}{ll}
    \pl_t\psi(t,x)=& [I_n-M(t,x;\psi(t,x))]^{-1}[\pl_tV(t,x;\l)](\l=\psi(t,x)) \\
   \pl_i\psi(t,x)=& [I_n-M(t,x;\psi(t,x))]^{-1}[\pl_iV(t,x;\l)](\l=\psi(t,x)),\,\,i\in\{1,...,n\}
  \end{array}
\right.
\end{equation}
Combining \eqref{eq..40} and \eqref{eq..41} we get the conclusion that the H-J equations \eqref{eq..35} are valid. In conclusion, the piecewise smooth scalar function
\begin{equation*}\label{eq..42}
\{u(t,x)=\widehat{u}(t;\psi(t,x)):t\in[t_j,t_{j+1}),\,x\in \hbox{int} B(x_*,\g),\,j\geq 0\}
\end{equation*}
fulfils the original H-J equations with jumps \eqref{eq:1}. In addition, we get
\begin{equation*}\label{eq..43}
u(t,x)\in[a,b]\,\,\hbox{and}\,\,\,|u(t,x)-u_*|\leq(\frac{1}{2})^j,\,\,\,t\in[t_j,t_{j+1}),\,\,j\geq0
\end{equation*}
for any $x\in B(x_*,\g)$ and as far as $\pl_iu(t,x)=<\pl_\l\widehat{u}(t;\psi(t,x)),\pl_i\psi(t,x)>$ we see easily that (see \eqref{eq;109} and \eqref{eq..41})
\begin{equation*}\label{eq..44}
|\pl_iu(t,x)|\leq L_i\frac{K_1}{1-\rho}(\frac{1}{2})^j,\,\,t\in[t_j,t_{j+1}),\,\,x\in B(x_*,\g),\,j\geq0
\end{equation*}
where $L_i=max\{|\pl_{x_i}\widehat{G}(\s)[x]|:\,\s\in[-\b,\b],\,x\in B(x_*,\g)\}$ for $i\in\{1,...,n\}$. The proof is complete.
\end{proof}
\noindent\textbf{COMMENT}\\
The result presented in Theorem \ref{th;1} relies essentially on the special structure we have assumed for the vector field $g(x,u)=\a(u)\widehat{g}(x),\,\widehat{g}\in\ml1(\mbn,\mbn),\,\,\a\in\ml1([a,b])$, where $a=u_*-1,b=u_*+1$ and $\a(u_*)=0$. A relaxation of this hypothesis allows vector fields
$$g(x,u)=\mathop{\sum}\limits_{k=1}^{l}\a_k(u)\widehat{g}_k(x),\,\a_k\in\ml1([a,b]),\,\,\widehat{g}_k\in\ml1(\mbn;\mbn)$$
where $\a_k(u_*)=0,\,\,k\in\{1,...,l\}$, for some $u_*\in\mb$. To get an asymptotic stable solution we must assume, in addition, that $\{\widehat{g}_1,...,\widehat{g}_l\}\subseteq\ml1(\mbn;\mbn)$ are commuting using the standard Lie product. The second theorem of this paper is encompassing the details we need in order to get an asymptotic stable solution when the vector field $g(x,u)=\a_1(u)\widehat{g}_1(x)+\a_2(u)\widehat{g}_2(x)$ is defined by $\a_k\in\ml1([a,b]),\,\,\widehat{g}_k\in\ml1(\mbn;\mbn),\,\,k=1,2$, where $\a_k(u_*)=0,\,\,\,k\in\{1,2\}$, for some $u_*\in\mb$ fixed. Everywhere in what follows, we take a domain $D=B(x_*,\g)\subset\mbn$, where $x_*\in\mbn$ and $\g>0$ are fixed. Denote the standard Lie product $[\widehat{g}_1,\widehat{g}_2]=\{\{\pl_x\widehat{g}_1\}\widehat{g}_2-\{\pl_x\widehat{g}_2\}\widehat{g}_1\},\,\,x\in\mbn$ and let $a=u_*-1,\,\,b=u_*+1$, where $u_*\in\mb$ is fixed. Using the hamiltonian function
$$H(x,u,p)=<p,g(x,u)>+L(u),\,\,g(x,u)=\a_1(u)\widehat{g}_1(x)+\a_2(u)\widehat{g}_2(x)$$
We assume that $\{\widehat{g}_1,\widehat{g}_2\}\subseteq\ml1(\mbn;\mbn)$ and $L\in\ml1(\mb)$ fulfil the following conditions
\begin{equation}\label{eq..45}
\left\{
  \begin{array}{ll}
   (a)\,\,\, L(u_*)=&0,\,\,\pl_u L(u)\geq0,\,\,u\in[a,b]\\
  (b)\,\,\,\a_k(u_*)=&0,\,\,k=1,2\\
(c)\,\,\,[\widehat{g}_1,\widehat{g}_2](x)&=0,\,\,\hbox{for any}\,\,x\in B(x_*,3\g)
  \end{array}
\right.
\end{equation}
The corresponding (H-J) equations with jumps are described by
\begin{equation*}\label{eq..46}
\left\{
  \begin{array}{ll}
\pl_tu+H(x,u,\pl_xu&)=0,t\in[t_j,t_{j+1}),\,x\in D=B(x_*,\g)\\
  u(t_j,x)&=u(t_{j^-},x)+h(u(t_{j^-},x))\Delta y(t_j),\,\,j\geq0,x\in D\\
u(0,x)&=u_*+u_0(x),\,\,x\in D
  \end{array}
\right.
\end{equation*}
where the scalar jump function $h\in\ml1([a,b])$ and
 \begin{equation*}\label{eq..47}
\Delta y(t_j)=y(t_j)-y(t_{j^-})\geq0\,\,\,\,\hbox{for any}\,\,j\geq1
\end{equation*}
are associated with the scalar piecewise constant process $\{y(t)=0\,:\,\,y(0)=0\}$.\\
Define
\begin{equation*}\label{eq..48}
\left\{
  \begin{array}{ll}
    C_1= & max\{|\pl_u\a_1(u)|+|\pl_u\a_2(u)|:\,\,u\in[a,b]\},\\
   C_2 =& max\{|\widehat{g}_1(x)|+|\widehat{g}_2|:\,\,x\in B(x_*,3\g)\}
  \end{array}
\right.
\end{equation*}
Take $u_0$ and $h$ such that
\begin{equation}\label{eq..49}
\left\{
  \begin{array}{ll}
    &(a)\,\,\,u_0\in \mlb1(\mbn),\,\,u_0(x_*)=0,\,\,sup|u_0(x)|=K_0<1\,\,\\
    &\,\,\,\,\,\,\,\,\,\,\,\,\hbox{and}\,\,sup|\pl_xu_0(x)|=K_1; \\
  &(b)\,\,\, h\in\ml1([a,b]),\,h(u_*)=0,\,\,\pl_uh(u)\leq\delta<0,\,\,u\in[a,b]\,; \\
&(c)\,\,\,2d C_1 C_2 K_1=\rho\in(0,\frac{1}{2}),\,\,\hbox{where}\,\,d=\mathop{max}\limits_{j\geq0}(t_{j+1})\,\\
&\,\,\,\,\,\,\,\,\,\hbox{and}\,\,\b=2dC_1\,\hbox{satisfies}\,\,\b C_2\leq\frac{\g}{2}
  \end{array}
\right.
\end{equation}
\begin{Theorem}\label{th;2}
Consider $H(x,u,p)=<p,g(x,u)>+L(u)$, where $g(x,u)=\a_1(u)\widehat{g}_1(x)+\a_2(u)\widehat{g}_2(x)$ and $L(u)$ fulfil \eqref{eq..45}. Let $u_0$ and $h$ be such that \eqref{eq..49} are satisfied. Take $\Delta y(t_j)\geq0$ verifying
\begin{equation}\label{eq..50}
-1\leq\pl_u h(u)\Delta y(t_j)\leq-\frac{1}{2},\,\,u\in[a,b],\,\,j\geq1
\end{equation}
Then there exists a global solution $\{u(t,x)\,:t\geq0,\,x\in D\}$ of the H-J equations with jumps \eqref{eq..38} which is asymptotically stable satisfying
\begin{equation}\label{eq..51}
\left\{
  \begin{array}{ll}
    |u(t,x)-u_*|\leq(\frac{1}{2})^j,\,\,t\in[t_j,t_{j+1}),\,\,x\in D,\,j\geq0 \\
  |\pl_iu(t,x)|\leq L_i\frac{K_1}{1-\rho}(\frac{1}{2})^j,\,\,t\in[t_j,t_{j+1}),\,\,x\in D,\,j\geq0 ,\,i\in\{1,...,n\}
  \end{array}
\right.
\end{equation}
for some constant $L_i>0$, where $K_1>0$ and $\rho\in(0,1)$ are given in \eqref{eq..49}
\end{Theorem}
\begin{proof}
We use the same arguments as in the proof of Theorem \ref{th;1}. By hypothesis, the conclusion of Lemma \ref{l;3} regarding the global solution $\{\widehat{u}(t;\l):\,t\geq0,\,\l\in\mbn\}$ are valid. The following equations are satisfied
\begin{equation*}\label{eq..52}
\left\{
  \begin{array}{ll}
   \frac{d\widehat{u}}{dt}=-L(\widehat{u}),\,\,\widehat{u}(t_{j^-};\l)+h(\widehat{u}(t_{j^-};\l))\Delta y(t_j),\, t\in[t_j,t_{j+1}),\,\,j\geq0\\
\widehat{u}(0;\l)=u_*+u_0(\l),\,\,\l\in\mbn
  \end{array}
\right.
\end{equation*}
\begin{equation*}\label{eq..53}
\left\{
  \begin{array}{ll}
   |\widehat{u}(t;\l)-u_*|\leq|u_0(\l)|\leq K_)<1,\,\,t\geq0,\,\l\in\mbn\\
|\pl_\l\widehat{u}(t;\l)|\leq|\pl_\l u_0(\l)|,\,\,t\geq0,\,\,\l\in\mbn
  \end{array}
\right.
\end{equation*}
In addition, using (\ref{eq..49},a,b) and \eqref{eq..50}, by a direct computation, we get
\begin{equation*}\label{eq..54}
\left\{
  \begin{array}{ll}
  |\pl_\l\widehat{u}(t;\l)|\leq(\frac{1}{2})^jK_1 ,\,\,t\in[t_j,t_{j+1}),\,\,j\geq0,\,\l\in\mbn\\
|\widehat{u}(t;\l)-u_*|\leq|\widehat{u}(t_j;\l)-u_*|\leq(\frac{1}{2})^j,\,\,t\in[t_j,t_{j+1}),\,\,j\geq0,\,\,\l\in\mbn
  \end{array}
\right.
\end{equation*}
This time, $\{\widehat{x}(t;\l)\,:t\geq0,\,\l\in B(x_*,2\g)\subseteq\mbn\}$ is the global solution of the characteristic system
\begin{equation*}\label{eq..55}
\left\{
  \begin{array}{ll}
  \frac{d\widehat{x}}{dt}(t;\l)=\a_1(\widehat{u}(t;\l))\widehat{g}_1(\widehat{x}(t;\l))+\a_2(\widehat{u}(t;\l))\widehat{g}_2(\widehat{x}(t;\l)),\,\,t\geq0\\
\widehat{x}(0;\l)=\l\in B(x_*,2\g)
  \end{array}
\right.
\end{equation*}
As far as $[\widehat{g}_1,\widehat{g}_2](x)=0,\,\,\forall\,\,x\in B(x_*,3\g)$, we may and do write $\widehat{x}(t;\l)$ as follows
\begin{equation}\label{eq..56}
\widehat{x}(t;\l)=\widehat{G}_1(\t_1(t,\l))\circ\widehat{G}_2(\t_2(t,\l))[\l],\,\,\,t\geq0,\,\,\,\l\in B(x_*,2\g)
\end{equation}
provided $\{\widehat{x}(t;\l)\in B(x_*,3\g):\,t\geq0,\,\,\l\in B(x_*,2\g)\}$. Here $\{\widehat{G}_k(t_k)[\l]:t_k\in[-\b,\b],\,\,\l\in B(x_*,2\g+\frac{\g}{2})\}$ is the local flow generated by $\widehat{g}_k\in\ml1(\mbn;\mbn)$
and using that
\begin{eqnarray*}\label{eq..57}
\t_k(t,\l)&=\int_{0}^{t}\a_k(\widehat{u}(s;\l))ds=\\ \nonumber
&=\int_{0}^{t}[\int_{0}^{1}\pl_u\a_k(u_*+\theta[\widehat{u}(s;\l)-u_*])d\theta](\widehat{u}(s;\l)-u_*)ds,\,k\in\{1,2\}
\end{eqnarray*}
we get (see \eqref{eq..54} and (\ref{eq..49},c))
\begin{equation*}\label{eq..58}
|\t_k(t,\l)|\leq C_1\int_{0}^{1}|\widehat{u}(s;\l)-u_*|ds\leq C_1\mathop{\sum}\limits_{j=0}^{\infty}\mathop{\int}\limits_{t_j}^{t_{j+1}}(\frac{1}{2})^jds\leq 2d C_1=\b
\end{equation*}
for any $t\geq 0,\,\,\l\in\mbn,\,\,k\in\{1,2\}$, where $\b C_2\leq\frac{\g}{2}$. It shows that the flow $\{\widehat{x}(t;\l):t\geq0,\,\l\in B(x_*,2\g)\}$ defined in \eqref{eq..56} fulfils $\widehat{x}(t,\l)\in B(x_*,3\g)$ and the equations
\begin{equation*}\label{eq..59}
x=\widehat{x}(t;\l)=\widehat{G}_1(-\t_1(t;\l))\circ\widehat{G}_2(-\t_2(t;\l))[x],\,\hbox{for}\,\,x\in B(x_*,\g)=D,\,\,\l\in B(x_*,2\g)
\end{equation*}
are equivalent with
\begin{equation}\label{eq..60}
\l=\widehat{G}_2(-\t_2(t;\l))\circ\widehat{G}_1(-\t_1(t;\l))[x],\,\hbox{for}\,\,\l\in B(x_*,2\g)\,\,\hbox{and}\,\,x\in D
\end{equation}
Denote
\begin{equation*}\label{eq..61}
V(t,x;\l)=\widehat{G}_2(-\t_2(t;\l))\circ\widehat{G}_1(-\t_1(t;\l))[x]=\widehat{G}_1(-\t_1(t;\l))\circ\widehat{G}_2(-\t_2(t;\l))[x]
\end{equation*}
for $t\geq0,\,\,x\in B(x_*,\g),\,\,\l\in B(x_*,2\g)$. To prove that the smooth mapping $V(t,x;\l)$ is a contraction with respect to $\l\in B(x_*,2\g)$ we compute
\begin{equation*}\label{eq..62}
M(t,x;\l)=\pl_\l V(t,x;\l)=-[\widehat{g}_1(V(t,x;\l))\pl_\l\t_1(t;\l)+\widehat{g}_2(V(t,x;\l))\pl_\l\t_2(t;\l)]
\end{equation*}
and using \eqref{eq..54} and (\ref{eq..49},c) we obtain
\begin{equation*}\label{eq..63}
|M(t,x;\l)|\leq\rho\in(0,\frac{1}{2}),\,\,\hbox{for any}\,\,t\geq0,\,x\in B(x_*,\g),\,\l\in B(x_*,2\g)
\end{equation*}
It allows to use the contractive mapping theorem for the integral equations \eqref{eq..60} and to get the smooth solution $\l=\psi(t,x)=\mathop{lim}\limits_{k\rightarrow\infty}\l_k(t,x)\in B(x_*,2\g)$ satisfying
\begin{equation*}\label{eq..64}
\left\{
  \begin{array}{ll}
  \l_0(t,x)&=x_*,\,\,|\l_k(t,x)-x_*|\leq\frac{1}{1-\rho}\g\leq2\g,\,\,\forall\,\,k\geq0, \\
   \psi(t,x)&=V(t,x;\psi(t,x)),\,\,t\geq0,\,\,x\in B(x_*,\g)
  \end{array}
\right.
\end{equation*}
Define $\{u(t,x)=\widehat{u}(t,\psi(t,x)):\,t\in[t_j,t_{j+1}),\,\,x\in B(x_*,\g),\,j\geq0\}$ and using the same argument as in Theorem \ref{th;1} we get the conclusions \eqref{eq..51} and the proof is complete.
\end{proof}

\begin{flushright}
\begin{tabular}{c}
  Abdus Salam School of Mathematical Science \\
  GC University, \\
  68 B, New MuslimTown, \\
  Lahore, Pakistan. \\
  54600 \\
\end{tabular}
\end{flushright}


\begin{thebibliography}{99}
\bibitem{01} B.Iftimie , I.Molnar, C.Vˆarsan, Solutions of some elliptic equations associated with a piecewise continuous process, Revue Roumaine Math Pures Appl.53(2008)No4, 323-338\\\\\\
\end{thebibliography}
\end{document}